\documentclass[preprint,12pt]{elsarticle}

\usepackage{amssymb}

\usepackage{dcolumn}
\usepackage{bm}
\usepackage{amsfonts}
\usepackage{hyperref}
\usepackage{color}

\usepackage{graphicx}
\usepackage{colordvi}
\usepackage{mathtools}
\usepackage{empheq}

\def\be{\begin{equation}}
\def\ee{\end{equation}}
\def\ba{\begin{eqnarray}}
\def\ea{\end{eqnarray}}

\journal{Physics of the Dark Universe}

\begin{document}

\begin{frontmatter}



\title{Isocurvature fluctuations in the effective Newton's constant}


\author{D. Paoletti}
\address{INAF/OAS Bologna, Osservatorio di Astrofisica e Scienza dello Spazio, \\
Area della ricerca CNR-INAF, via Gobetti 101, I-40129 Bologna, Italy}
\address{INFN, Sezione di Bologna, \\
via Irnerio 46, I-40126 Bologna, Italy}
\author{M. Braglia} 
\address{DIFA, Dipartimento di Fisica e Astronomia,\\
Alma Mater Studiorum, Universit\`a degli Studi di Bologna,\\
via del Battiferro, I-40129 Bologna, Italy}
\address{INAF/OAS Bologna, Osservatorio di Astrofisica e Scienza dello Spazio, \\
Area della ricerca CNR-INAF, via Gobetti 101, I-40129 Bologna, Italy}
\address{INFN, Sezione di Bologna, \\
via Irnerio 46, I-40126 Bologna, Italy}
\author{F. Finelli} 
\address{INAF/OAS Bologna, Osservatorio di Astrofisica e Scienza dello Spazio, \\
Area della ricerca CNR-INAF, via Gobetti 101, I-40129 Bologna, Italy}
\address{INFN, Sezione di Bologna, \\
via Irnerio 46, I-40126 Bologna, Italy}
\author{M. Ballardini}
\address{Department of Physics \& Astronomy, \\ 
University of the Western Cape, Cape Town 7535, South Africa}
\address{INAF/OAS Bologna, Osservatorio di Astrofisica e Scienza dello Spazio, \\
Area della ricerca CNR-INAF, via Gobetti 101, I-40129 Bologna, Italy}
\author{C. Umilt\`a,} 
\address{Department of Physics, \\ 
University of Cincinnati, Cincinnati, Ohio, 45221, USA}

\begin{abstract}
We present a new isocurvature mode present in scalar-tensor theories of gravity that corresponds to a regular growing solution in which 
the energy of the relativistic degrees of freedom and the scalar field that regulates the gravitational strength compensate 
during the radiation dominated epoch on  scales much larger than the Hubble radius. We study this isocurvature mode and its impact on 
anisotropies of the cosmic microwave background for the simplest scalar-tensor 
theory, i.e. the extended Jordan-Brans-Dicke gravity, in which the scalar field also drives the acceleration of the Universe.
We use {\em Planck} data to constrain the amplitude of this isocurvature mode in the case of fixed correlation with the adiabatic mode and we show 
how this mode could be generated in a simple two field inflation model.
\end{abstract}


\end{frontmatter}

\section{Introduction}

 Since its discovery with the analysis of SNIa light curve of the Supernova Cosmology Project 
\cite{Perlmutter:1998np} and High-Z Supernova Search Team \cite{Riess:1998cb},
the acceleration of the Universe at $z \sim 1$ has been confirmed by a host of cosmological observations in the last 20 years.
A cosmological constant $\Lambda$, which is at the core of the minimal concordance cosmological $\Lambda$CDM model in agreement 
with observations, is the simplest explanation of the recent acceleration of the Universe, but several alternatives 
have been proposed either replacing $\Lambda$ by a dynamical component or modifying Einstein gravity (see \cite{Sahni:1999gb,Peebles:2002gy,Clifton:2011jh} 
for reviews on dark energy/modified gravity). 

If a dynamical component as quintessence, which varies in time and space, drives the Universe into acceleration instead of 
$\Lambda$ \cite{Ratra:1987rm,Caldwell:1997ii}, not only the homogeneous cosmology is modified, also its fluctuations cannot be neglected and their 
behaviour can help in distinguishing structure formation in different theoretical models. 
This dynamical component can, in combination with the other cosmic fluids (radiation, baryons, cold dark matter, neutrinos), 
lead not only to adiabatic curvature  perturbations, but to a mixture which includes an isocurvature component. 
Isocurvature perturbations appear when the relative energy density and pressure perturbations 
of the different fluid species compensate to leave the overall curvature perturbations unchanged for scales much larger than the Hubble radius.

In the case of quintessence, it was found that its fluctuations are very close to be adiabatic during
a tracking regime in which the parameter of state of quintessence mimics the one of the component dominating 
the total energy density of the Universe \cite{Abramo:2001mv}.
In the case of thawing quintessence models, in which a tracking regime is absent, isocurvature
quintessence fluctuations are instead allowed \cite{Abramo:2001mv,Moroi:2003pq}. From the phenomenological point of view, a mixture of curvature 
and quintessence isocurvature perturbations is an interesting explanation of the low amplitude of the quadrupole 
and more in general of the low-$\ell$ anomaly of the cosmic microwave background (CMB henceforth) anisotropies 
pattern \cite{Moroi:2003pq,Gordon:2004ez}. 

In this paper we study a new isocurvature mode which is present in scalar-tensor 
theories of gravity,
in which the scalar field responsible for the acceleration of the Universe also regulates 
the gravitational strength through its non minimal coupling to gravity 
\cite{Perrotta:1999am,Chiba:1999wt,Uzan:1999ch,Bartolo:1999sq,Boisseau:2000pr,Baccigalupi:2000je}. 
These models are also known as extended quintessence \cite{Perrotta:1999am}. 
We will study the effect of this new isocurvature mode on CMB anisotropies and show that this can be generically excited 
during inflation with an amplitude allowed by Planck data.

\section{The model}

\begin{figure}
\includegraphics[width=0.5\columnwidth]{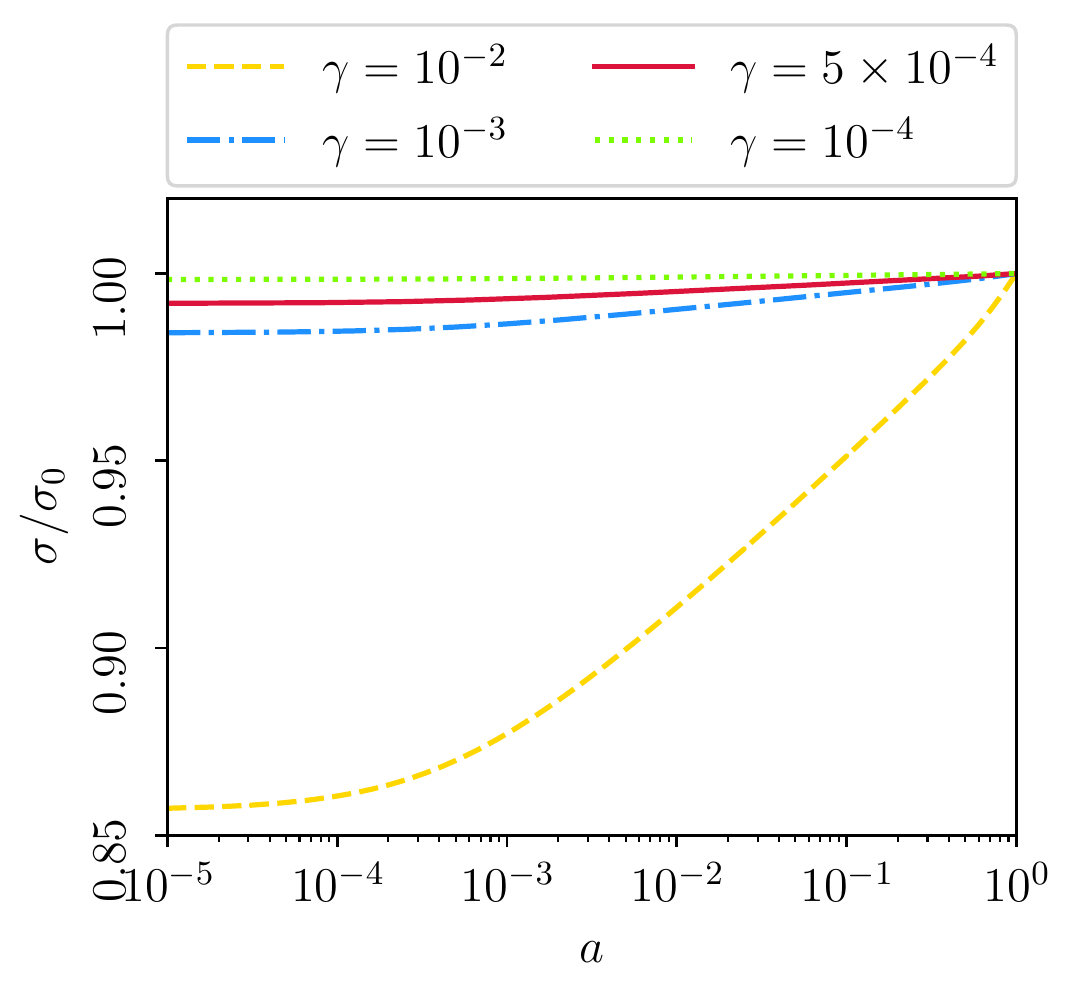} \includegraphics[width=0.5\columnwidth]{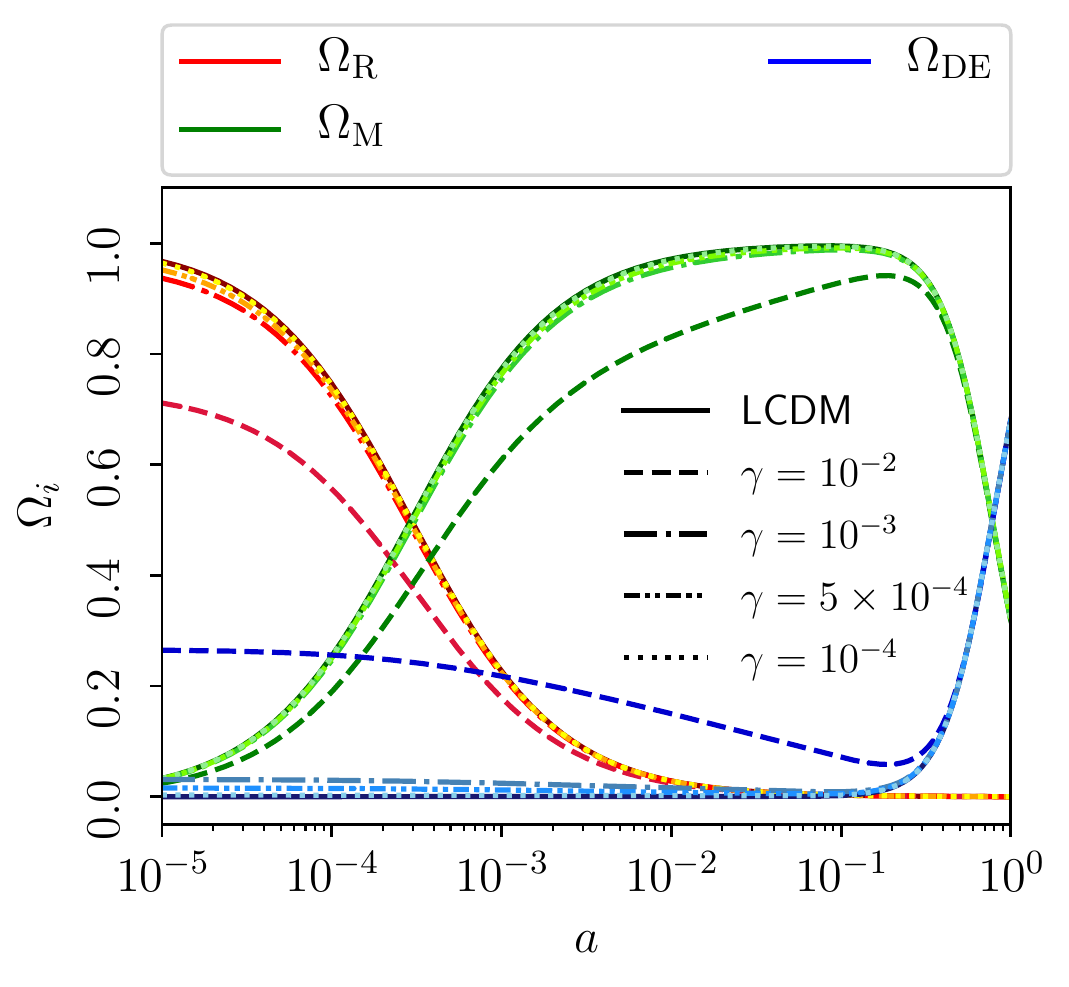} 
\caption{Example of an evolution for $\sigma/\sigma_0$ (left panel) and $\Omega_i$ (right panel) as function of $a$
for different choices of $\gamma$ for a quartic potential. The value $\sigma_0$ of the
scalar field at present is fixed consistently with the Cavendish-type measurement of the gravitational constant
$G = 6.67\times10^{-8}$ cm$^3$ g$^{-1}$ s$^{-2}$: $\gamma \sigma_0^2 = \frac{1}{8\pi G}\frac{1 + 8\gamma}{1 + 6\gamma}$. 
}
\label{phiomega}
\end{figure}

We consider the simplest scalar-tensor gravity 
theory, in the original Jordan frame, describing the late time Universe:
\begin{equation}
\!\!\!\!\!\!\!\!\!\!\!\!\!\!S = \int d^4x \sqrt{-g}\, \Bigl[ \frac{\gamma \sigma^2 R}{2} - \frac{g^{\mu \nu}}{2}
\partial_{\mu} \sigma \partial_{\nu} \sigma - V(\sigma) + \mathcal{L}_m \Bigr] 
\label{model}
\end{equation}
where $\mathcal{L}_m$ denotes the matter content (baryon, CDM, photons, neutrinos), 
$\sigma$ is the Jordan-Brans-Dicke (JBD) scalar field whose equation of motion is:
\begin{equation}
\!\!\!\!\!\!\!\!\!\!\!\!\!\!\ddot \sigma + 3 H \dot \sigma + \frac{\dot \sigma^2}{\sigma} + \frac{\sigma^4}{(1+6\gamma)} \Big(\frac{V}{\sigma^4}\Big)_{,\sigma}=\frac{\sum_i(\rho_i -3p_i)}{(1+6\gamma)\sigma}
\label{KG}
\end{equation}
Note that the above 
induced gravity Lagrangian can 
be recast in a extended JBD theory of gravity \cite{JBD} by a redefinition 
of the scalar field $\phi = \gamma \sigma^2$ and $\omega_{\mathrm BD}=(4 \gamma)^{-1}$. We will consider the case of a
non-tracking potential as $V(\sigma)\propto\sigma^4$ \cite{Cooper:1982du,Wetterich:1987fm,Finelli:2007wb} in the following (see \cite{Ballardini:2016cvy} 
for other monomial potentials).
The background cosmological evolution is displayed in Fig. \ref{phiomega}:
deep in the radiation era, $\sigma$ is \emph{nearly frozen} as demonstrated by analytic methods \cite{Gurevich,Finelli:2007wb}; 
during the subsequent matter dominated era, $\sigma$ is driven by non-relativistic 
matter to larger values.
These two stages of evolution are quite model independent for $\sigma$ with a very small effective mass and 
a coupling $\gamma\sigma^2 R$: 
the potential $V(\sigma)$ kicks in only in a third stage at recent times determining the rate of the accelerated 
expansion \cite{Barrow:1990nv,Cerioni:2009kn} (see Fig. \ref{phiomega}).
Following Ref. \cite{Boisseau:2000pr}, we also plot the effective energy density
fractions, as defined in Eq. (2.5) of Ref. \cite{Umilta:2015cta}, in the right panel of Fig. \ref{phiomega}.

The evolution of linear perturbations in the adiabatic initial conditions has been considered for the most recent constraints 
on this class of scalar-tensor theories \cite{Umilta:2015cta,Ballardini:2016cvy}. The so-called adiabatic initial 
condition \cite{Ma:1995ey} are regular solution to the  Boltzmann, Klein-Gordon and Einstein equations 
in scalar-tensor gravity characterized by a constant curvature perturbation for scales much larger the Hubble radius during the radiation dominated
epoch. 

In this paper we wish to present the original result for a more general 
initial condition which include a mixture of the adiabatic and a new isocurvature solution between the 
relativistic degrees of freedom and the scalar 
field. The latter is a new solution which is obviously absent in $\Lambda$CDM and is an example of the 
generic new independent growing solution 
within scalar-tensor theories of gravity. 

\section{The initial conditions} 

In the following we use the synchronous gauge for metric fluctuations (see Eqs. (1-4) of Ref. \cite{Ma:1995ey}) and we denote by 
$\delta_i$ the energy density contrasts, $\theta_i=\imath k_j v_i^j$ the velocity potentials  and  $\sigma_\nu$ the neutrino anisotropic pressure.
The indices $i=b,\,c,\,\gamma,\,\nu$ denote baryons, CDM, photons, and neutrinos respectively.
The scalar field fluctuation is $\delta \sigma$.
The perturbed Einstein equations are given by:
\begin{align}
k^2 \eta-\frac{1}{2}\mathcal{H}h'&=-\frac{a^2 \delta\tilde{\rho}}{2}\,,\\
k^2 \eta'&= \frac{a^2(\tilde{\rho}+\tilde{P})\tilde{\theta}}{2}\,,\\
h''+2\mathcal{H}h'-2k^2\eta&=-3 a^2 \delta\tilde{P} \,, \\
(h''+6\eta'')+2\mathcal{H}(h'+6\eta')-2k^2 \eta&=-3 a^2(\tilde{\rho}+\tilde{P})\tilde{\sigma},
\end{align}
where
\begin{align}
\delta\tilde{\rho}\equiv&\frac{\delta\rho_m}{\gamma\sigma^2}+\frac{h'\sigma'}{a^2 \sigma}
-\frac{2}{a^2}\Biggl\{\frac{\delta\sigma'}{\sigma}\left(\mathcal{H}-\frac{\sigma'}{2\gamma\sigma}\right)+\notag\\
&+\frac{\delta\sigma}{\sigma}\left[\frac{a^2\rho_m}{\gamma\sigma^2}+\frac{{\sigma'}^2}{2\gamma\sigma^2}
+\frac{a^2}{\gamma\sigma}\left(\frac{V}{\sigma}-\frac{V,_{\sigma}}{2}\right)-\frac{3\mathcal{H}\sigma'}{\sigma}+k^2\right]  \Biggr\}, \notag
\end{align}
\begin{equation}
(\tilde{\rho}+\tilde{P})\tilde{\theta}\equiv \frac{(\rho_m+P_m)\theta_m}{\gamma\sigma^2}
+\frac{2 k^2}{a^2}\left\{\frac{\delta\sigma}{\sigma}\left[\frac{\sigma'}{2\gamma\sigma}(1+2\gamma)
-\mathcal{H}\right]+\frac{\delta\sigma'}{\sigma}\right\},\notag
\end{equation}
\begin{align}
\delta\tilde{P}\equiv&\frac{1}{(1+6 \gamma)\sigma^2}\left(2\delta\rho_m+\frac{\delta P_m}{\gamma}\right)
-\frac{1}{3 a^2}\Biggl\{\frac{3\delta\sigma'}{\sigma} \left(2\mathcal{H}-\frac{\sigma'}{\gamma\sigma}\right)+\notag\\
&+\frac{\delta\sigma}{\sigma}\Biggl[\frac{6 a^2 P_m}{\gamma\sigma^2}+\frac{12 a^2 (\rho_m-3 P_m)}{(1+6\gamma)\sigma^2}
+\frac{3\sigma'}{\sigma}\left(\frac{\sigma'}{\gamma\sigma}-2\mathcal{H}\right)
+2k^2+\notag\\
&+\frac{6a^2 }{(1+6\gamma)}\left(V,_{\sigma\sigma}+\frac{V,_{\sigma}}{2\gamma\sigma}(1-4\gamma)-\frac{V}{\gamma\sigma^2}(1-2\gamma) \right)     \Biggr]         +\frac{h'\sigma'}{\sigma}      \Biggr\},\notag
\end{align}
\begin{equation}
(\tilde{\rho}+\tilde{P})\tilde{\sigma}\equiv \frac{(\rho_m+P_m)\sigma_m}{\gamma\sigma^2}+\frac{1}{3 a^2}\left[\frac{4 k^2\delta\sigma}{\sigma}
+2(h'+6\eta')\frac{\sigma'}{\sigma}\right].
\end{equation}
Also, the perturbed Klein-Gordon equation is:
\begin{multline}
\label{kleno}
\delta\sigma''=-2\delta\sigma'\left(\mathcal{H}+\frac{\sigma'}{\sigma}\right)-\delta\sigma\Biggl[k^2-\frac{{\sigma'}^2}{\sigma^2}
+\frac{a^2(\rho_m-3P_m)}{(1+6\gamma)\sigma^2}+\\
+\frac{a^2}{(1+6\gamma)}\left(V,_{\sigma\sigma}+\frac{4 V}{\sigma^2}-\frac{4 V,_{\sigma}}{\sigma}\right)
+\frac{a^2(\delta\rho_m-3\delta P_m)}{(1+6\gamma)\sigma}-\frac{h'\sigma'}{2}   \Biggr].
\end{multline}

The adiabatic plus the new isocurvature solution, in the background considered, is given by:

\begin{eqnarray}
	\delta_{\gamma}&=& \delta_\nu = C \left[-\frac{2}{3}  k^2 \tau^2+\frac{2 \omega}{15}  k^2 \tau^3 \right] + D \Big[-1 - \frac {2 \omega } {3} \tau + \frac { 3 (15\gamma + 2)\omega^2 + 4 k^2 } {24}  \tau^2    \Big]\,,\nonumber\\
	\theta_{\gamma}&=&C \Big[-\frac{k^4 \tau^3}{36} + \frac{\omega  (5 (1+6 \gamma ) {R_b}+R_{\gamma})}{240 R_{\gamma}}k^4 \tau^4]\nonumber\\
	&&
	+D\Big[ -\frac{k^2}{4} \tau+ \frac{\omega}{48} \left (\frac {9 (1+6\gamma )  R_b} {R_{\gamma}} - 4  \right) k^2 \tau^2 \Big]\,,\nonumber\\	
	\delta_b&=& C \Big[ -\frac{k^2}{2} \tau^2+\frac{\omega}{10} k^2 \tau^3 \Big] +D\Big[ -\frac { \omega} {2} \tau + \frac {1} {8}\left (\frac {3 (15\gamma +2)\omega^2} {4 k} + k \right) k \tau^2   \Big]\,,\nonumber\\
	\delta_c&=& C\Big[ -\frac{k^2}{2} \tau^2+\frac{\omega}{10}  k^2 \tau^3 \Big]+D\Big[ -\frac {1} {2} \omega\tau 
+ \frac {3} {32} (15\gamma + 2) \omega^2 \tau^2  \Big]\,,\nonumber\\
	\delta_{\nu}&=& C\Big[-\frac{2}{3}  k^2 \tau^2+\frac{2}{15}  k^2 \tau^3 \omega\Big]+D\Big[ -1 - \frac {2 \omega } {3} \tau + \frac { 3 (15\gamma + 2)\omega^2 + 4 k^2 } {24}  \tau^2 \Big]\,,\nonumber\\
	\theta_{\nu}&=&C\Big[ -\frac{  (4 R_{\nu}+23)}{18 (4 R_{\nu}+15)}k^4 \tau^3+\frac{ \omega  \left(8 R^2_{\nu}+60 \gamma  (5-4 R_{\nu})+50 {R_{\nu}}+275\right)}{120 (2 R_{\nu}+15) (4 R_{\nu}+15)} k^4 \tau^4\Big] \nonumber\\
	&&+D\Big[ -\frac {1 } {4} k^2 \tau - \frac {1} {12} \omega k^2 \tau^2\Big]\,,\nonumber\\
	\sigma_{\nu}&=&C\Big[\frac{4  k^2 \tau^2}{3 (4  R_{\nu}+15)}+\frac{   (1+6 \gamma )  (4 R_{\nu}-5)  \omega k^2 \tau^3}{3 (4 R_{\nu}+15) (2 R_{\nu}+15)}\Big]\nonumber\\
	&&+D\Big[ \frac { k^2 \tau^2} {6 (4 R_{\nu} +15)}  - \frac {2\omega  (1+6\gamma )  (R_{\nu} + 5) k^2 \tau^3 } {3 (2 R_{\nu} +15) (4 R_{\nu} + 15)}\Big]\,,\nonumber\\
h&=&C \Big[ k^2 \tau^2-\frac{1}{5}   \omega k^2 \tau^3\Big]+D\Big[ \omega \tau - \frac {3} {16} (15\gamma + 2) \omega^2  \tau^2 \Big]\,,\nonumber
\end{eqnarray}
\begin{eqnarray}
	\eta&=& C\Big[ 2-\frac{ (4 R_{\nu}+5)}{6 (4 {R_{\nu}}+15)} k^2 \tau^2	\Big] \nonumber \\
&& +D\Big[ -\frac { \omega} {6} \tau + \frac {16 k^2 (R_{\nu} + 5) + 3 (15\gamma + 	2) (4 R_{\nu} + 15)\omega^2 } {96 (4 R_{\nu} + 15)} \tau^2  \Big]\,\nonumber\\
\frac{\delta \sigma}{\sigma_i} &=&C\Big[-\frac{1}{4} \gamma \omega k^2\tau^3+\frac{\gamma  \omega^2}{40} 
(4+15 \gamma ) k^2 \tau^4
	\Big]+D \Big[ -\frac{1} {2} + \frac {3} {4}\gamma \omega\tau \Big] \,, 
\label{deltasigmainitial}
\end{eqnarray}
where $\omega\equiv\frac{\rho_{\textup{mat}0}}{\sqrt{3\gamma\rho_{\textup{rad}0}}(1+6\gamma)\sigma_i}$ and $\sigma_i$ is the value of $\sigma$ deep in the radiation era.
In the above equations $C$ ($D$) encodes the primordial power spectrum for curvature (isocurvature) perturbations.

This new mode is present in scalar-tensor theories of gravity and corresponds to a regular growing solution in which
the energy densities of the relativistic degrees of freedom and the scalar field compensate
at leading order on scales much larger than the Hubble radius, as can be seen by inserting the new solutions 
for $\delta_\gamma \,, \delta_\nu$ and $\delta \sigma$ in Eq. (3).
This new mode gives a vanishing contribution to the gauge-invariant curvature perturbation in the comoving gauge $\mathcal{R}$ \cite{Lyth:1984gv} 
and therefore can be accounted as an isocurvature.
To complete the characterization of this new isocurvature mode as done 
for Einstein gravity \cite{BMT}, we note that the Newtonian potentials, as defined in \cite{Ma:1995ey}, are given  at leading order by $
\Psi=-\frac {  (R_{\nu} + 5)} {(4 R_{\nu} +
        15)} $ and
$
\Phi=\frac {2  (R_{\nu} + 5)} {(4 R_{\nu} +
        15)}$. 

\section{Impact on CMB anisotropies}
\begin{figure}
\begin{tabular}{c}
\includegraphics[width=\textwidth]{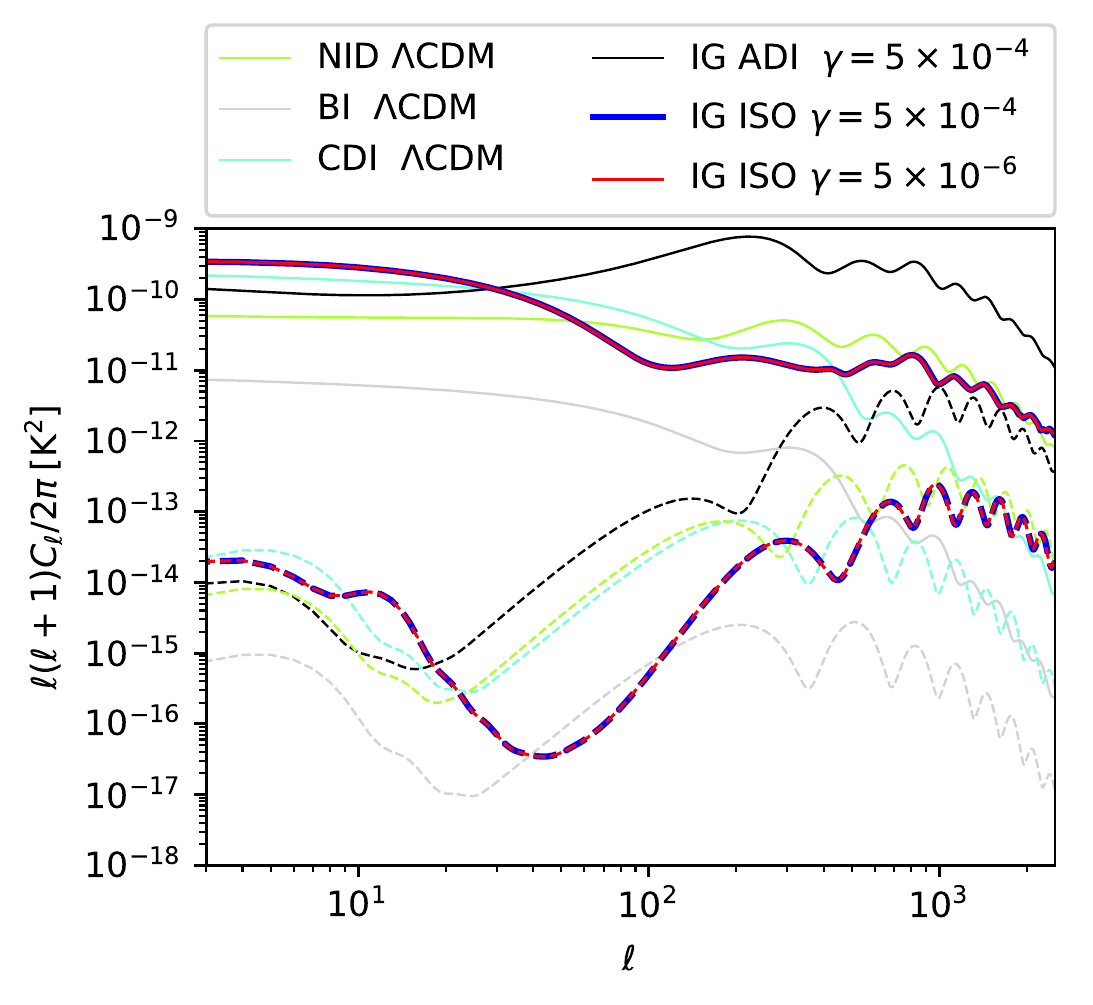}
\end{tabular}
\caption{CMB anisotropy angular power spectra in temperature, solid lines, and E-mode polarization, dashed lines. The black curve is the adiabatic case, thin curves represent the three standard $\Lambda$CDM isocurvature modes and the thick curve represent the new isocurvature mode. In order to compare the spectrum shapes we have assumed equal amplitude between isocurvatures and adiabatic mode, i.e. $f_{\mathrm{ISO}}=1$. Note that blue and red lines are superimposed.}
\label{Cl}
\end{figure}

In order to derive the predictions for the CMB anisotropy angular power spectra we have used an extension 
to the publicly available Einstein-Boltzmann code CLASS \footnote{\href{url}{www.class-code.net}} 
\cite{Lesgourgues:2011re,Blas:2011rf}, called CLASSig \cite{Umilta:2015cta}. 
CLASSig has been developed to derive the predictions for cosmological observables in induced gravity, and more in general 
scalar-tensor theories, solving for the perturbations but also for the background in order to 
derive the initial scalar field parameters which provide the cosmology in agreement with the measurements of the 
gravitational constant in laboratory Cavendish-type experiments.
CLASSig has been modified to include the initial conditions for the new isocurvature mode presented. 
In Fig. \ref{Cl} we show the comparison of the new mode with the adiabatic and standard 
isocurvature modes in the $\Lambda$CDM model within Einstein gravity. Fig. \ref{Cl} also shows the weak dependence of the new isocurvature mode on $\gamma$ at least for the small values consistent with the current cosmological 95\%CL upper bound $\gamma \lesssim 0.75 \times 10^{-3}$ \cite{Ballardini:2016cvy} and for Solar System constraints $\gamma \lesssim 0.6 \times 10^{-5}$ \cite{Bertotti:2003rm} .

\begin{figure}
\begin{tabular}{c}
\includegraphics[width=\textwidth]{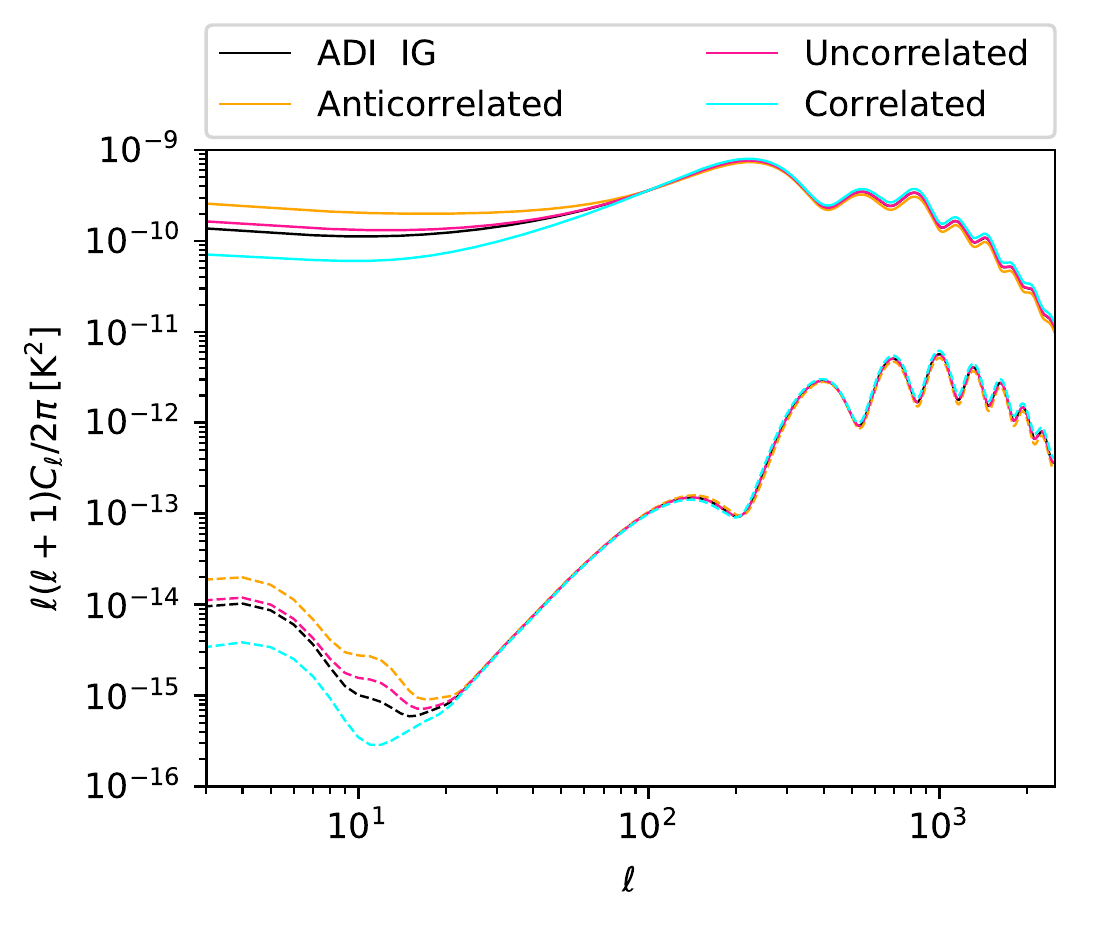}
\end{tabular}
\caption{CMB anisotropy angular power spectra in temperature, solid lines, and E-mode polarization, dashed lines, for the three different possibility of correlation between isocurvature and adiabatic modes with $f_{\textup{ISO}}=0.5$ and $n_{\textup{ISO}}=n_{\textup{ADI}}$.}
\label{ClC}
\end{figure}

The impact of a mixture of curvature and the new isocurvature initial conditions admitting a non-vanishing correlation $\theta$
on CMB anisotropies defined as \cite{Amendola:2001ni}
$C_\ell = C_\ell^{\mathrm {ADI}} + f_{\mathrm{ISO}}^2 C_\ell^{\mathrm {ISO}}+ 2 f_{\mathrm{ISO}} \cos\theta\, C_\ell^{\mathrm {CORR}}$
is displayed in Fig. \ref{ClC} ($f_{{\mathrm{ISO}}}$ being the relative fraction of isocurvature). Overall, 
the effect of this new isocurvature mode seems larger and on a wider range of multipoles than the quintessence 
isocurvature mode in Einstein gravity studied in \cite{Moroi:2003pq}.

\section{Comparison with data}. We now present the constraints on the new isocurvature amplitude with Planck data.
Since the effect of isocurvature perturbations
on the CMB anisotropy power spectra does not depend significantly on $\gamma$ for $\gamma \lesssim 10^{-3}$ (see Fig. \ref{Cl}), 
we fix $\gamma=5 \times 10^{-4}$ to contain the computational cost of our investigation. Such a value is either compatible with current cosmological observations
\cite{Umilta:2015cta,Ballardini:2016cvy} and conservatively close to the values tested 
in the comparison of Einstein-Boltzmann codes dedicated to JBD theories reported in \cite{Bellini:2017avd}.
We consider separately the three cases of correlation between adiabatic and isocurvature perturbation 
$\cos \theta = -1, 0, 1$ as in \cite{Ade:2015lrj}.
As data we consider Planck 2015 high-$\ell$ temperature likelihood at $\ell \ge 30$
in combination with the joint temperature-polarization likelihood for $2 \le \ell < 30$ and 
the Planck lensing likelihood \cite{Aghanim:2015xee,Ade:2015zua}.
To speed up the likelihood evaluation we use the foreground marginalized PlikLite likelihood at high $\ell$
which has been shown in \cite{Aghanim:2015xee} to be in good agreement with minimal extensions
of the $\Lambda$CDM model within Einstein gravity. We have also explicitly checked that we can reproduce with PlikLite
the constraints obtained with the full Planck high-$\ell$ temperature likelihood in \cite{Ballardini:2016cvy}.

The modified CLASSig which includes the general initial conditions
has been connected to the publicly available code
{\sc Monte Python}\footnote{\href{url}{https://github.com/brinckmann/montepython$\_$public}} \cite{Audren:2012wb,Brinckmann:2018cvx}
to compute 
the Bayesian probability distribution of cosmological parameters.
We vary the baryon density $\omega_{b}=\Omega_{b} h^2$, the cold dark matter density $\omega_{c}= \Omega_{c}h^2$ (with $h$ being
$H_0/100 {\rm km}\,{\rm s}^{-1}{\rm Mpc}^{-1}$), the reionisation optical depth $\tau_{\mathrm opt}$, 
the Hubble parameter $H_0$ at present time, $\ln ( 10^{10} A_S )$, $n_S$ and the isocurvature fraction $f_{\mathrm{ISO}}$ with a flat prior [0,0.8].
We sample the posterior using the Metropolis-Hastings algorithm \cite{Hastings:1970aa}  and imposing a Gelman-Rubin convergence
criterion \cite{Gelman:1992zz} of $R-1 < 0.02$.

We find no evidence at a statistical significant level for this new isocurvature mode. 
The 95\% CL bounds from the MCMC exploration are $f_{\textup{ISO}}<0.07$ for the fully anti-correlated case $\cos \theta = -1$, 
$f_{\textup{ISO}}<0.12$ for the fully correlated case $\cos \theta = 1$ and $f_{\textup{ISO}}<0.31$ for the uncorrelated 
case $\theta = \pi / 2$. 
These allowed abundances are slightly larger than those of the known isocurvature modes in Einstein gravity \footnote{Note that 
the relation $\beta=\frac{f_{\mathrm{ISO}}^2}{1+f_{\mathrm{ISO}}^2}$ between the isocurvature fraction $\beta$ in \cite{Ade:2015lrj} and 
$f_{\mathrm{ISO}}$ holds.}, although scale similarly with the degree of correlation \cite{Ade:2015lrj}.

\section{Isocurvature perturbations in the effective Newton's constant from inflation.} 
We now show that the amplitude of the new isocurvature mode in $\delta \sigma$
compatible with current data 
could be easily obtained in minimal inflationary
models within scalar-tensor gravity. We consider a two-field dynamics in which the scalar field $\sigma\equiv\Phi_2$ responsible for the late-time acceleration
was present during the inflationary stage driven by the inflaton $\phi\equiv\Phi_1$:
\begin{equation}
\!\!\!\!\!\!\!\!\!\!\!\!\!\!S = \int d^4x \sqrt{-g}\, \Bigl[ \frac{\gamma \sigma^2 R}{2} - \frac{g^{\mu \nu}}{2}
\partial_{\mu} \Phi_i \partial_{\nu} \Phi_i - V(\phi,\sigma)\Bigr] \,,
\label{eqn:IGinflation}
\end{equation}
where the indices $i=1,2$ are meant to be summed and $V(\phi,\sigma)=V_\textup{infl}(\phi)+V(\sigma)$.

Since $V(\sigma_0)/(3 \gamma \sigma_0^2 H_0^2) \sim 0.7$, $\sigma$ is effectively massless during inflation.
We assume that after inflation $\phi$ decays in ordinary matter and dark matter, which are coupled to $\sigma$ only gravitationally through
the term $\gamma R \sigma^2$. Once the Universe is thermalized, the evolution during the radiation and matter dominated era matches with what previously 
described for the background and perturbations: indeed, isocurvature perturbations in $\sigma$ 
are nearly decoupled from curvature perturbations during the radiation dominated period in which $\sigma$ is frozen. 

The two field dynamics in Eq. (\ref{eqn:IGinflation}) and the corresponding generation of curvature and isocurvature fluctuations
have been previously studied \cite{Starobinsky:1994mh,GarciaBellido:1995fz,Starobinsky:2001xq,DiMarco:2002eb,DiMarco:2005nq} either in the original Jordan frame or in the 
mathematically equivalent Einstein frame. Since in general curvature and isocurvature perturbations are not invariant under conformal transformations \cite{White:2012ya}, we work within the original frame in Eq. \eqref{model} consistently with the late time cosmology previously described. Under the assumption that $\sigma$ is 
subdominant during inflation, we find to lowest order in $\gamma$ and in the slow-roll parameters the isocurvature fraction on scale much larger than the Hubble radius \cite{Bragliaetal}:
\begin{equation}
\label{fisoinfl}
\frac{\mathcal{P}_{\mathcal{S}}(k_*)}{\mathcal{P}_{\mathcal{R}}(k_*)} \simeq \sin^2 \Delta \,
e^{\left(n_\mathrm{s}-n_\mathrm{iso}\right)(N-N_*)},
\end{equation}
where $n_\mathrm{s}$ ($n_\mathrm{iso}$) is the tilt of curvature (isocurvature) perturbations, 
$N_*$ is the number of $e$-folds to the end of inflation of the pivot scale $k_*$ and $\sin^2 \Delta$ is the 
isocurvature relative contribution at Hubble crossing.  
By considering the scalar 
tilt consistent with Planck data we use ($n_S=0.968\pm0.06$ at 68\% CL \cite{Ade:2015lrj}) 
and $[40,\,70]$ as a range for $N_*$, we find that the isocurvature fraction at the end 
of inflation in Eq. \eqref{fisoinfl} can be of the same order of magnitude of the Planck upper bound we obtain.

\section{Conclusions}
On concluding, we have presented a new growing independent solution 
in scalar-tensor gravity corresponding to an isocurvature mode between the scalar field which determines the evolution 
of the effective Newton's constant and the relativistic degrees of freedom. We have constrained  with Planck data
this new isocurvature mode when the scalar field is also responsible for the recent acceleration of the Universe  and we have shown how this mode can be generated during inflation. It will be interesting to see how the most recent CMB polarization data can further constrain this phenomenological aspect of scalar tensor theories. Work in this direction is in progress.


\section{Acknowledgements.}

We acknowledge useful discussions with R. Brandenberger, A. Starobinsky and S. Tsujikawa. 
We acknowledge support by the "ASI/INAF Agreement 2014-024-R.0 for the Planck LFI Activity of Phase
E2''. We also acknowledge financial contribution from the agreement ASI n. I/023/12/0 "Attivit\'a relative 
alla fase B2/C per la missione Euclid". DP and FF acknowledge financial support by ASI Grant 2016-24-H.0.
MB is supported by the South African SKA Project and by the Claude Leon Foundation.

\end{document}